\documentclass[a4paper,fleqn]{cas-sc}

\usepackage{CJKutf8}
\usepackage{amsmath}
\usepackage{url}
\usepackage{listings}
\usepackage{subfig}
\usepackage[capitalise]{cleveref}
\usepackage{natbib} 
\setcitestyle{square,numbers,sort&compress}
\usepackage{tabularx} 
\usepackage{array}
\usepackage{fontawesome} 
\usepackage{lettrine} 
\usepackage{fancyhdr} 
\usepackage{poemscol} 
\usepackage{lineno}
\input{Zallman.fd}

\definecolor{codegreen}{rgb}{0,0.6,0}
\definecolor{codegray}{rgb}{0.5,0.5,0.5}
\definecolor{codepurple}{rgb}{0.58,0,0.82}
\definecolor{backcolour}{rgb}{0.95,0.95,0.92}

\lstdefinestyle{mystyle}{
    backgroundcolor=\color{backcolour},   
    commentstyle=\color{codegray},
    keywordstyle=\color{codegreen},
    numberstyle=\tiny\color{codegray},
    stringstyle=\color{orange},
    basicstyle=\ttfamily\footnotesize,
    breakatwhitespace=false,         
    breaklines=true,                 
    captionpos=b,                    
    keepspaces=true,                 
    numbers=left,                    
    numbersep=5pt,                  
    showspaces=false,                
    showstringspaces=false,
    showtabs=false,                  
    tabsize=2
}

\usepackage{fancyhdr}  

\begin{document}


\let\WriteBookmarks\relax
\def\floatpagepagefraction{1}
\def\textpagefraction{.001}


\author[loyola]{R. Abbasi}
\author[zeuthen]{M. Ackermann}
\author[christchurch]{J. Adams}
\author[madisonpac]{S. K. Agarwalla\fnref{india}}
\author[brusselslibre]{J. A. Aguilar}
\author[copenhagen]{M. Ahlers}
\author[dortmund]{J.M. Alameddine}
\author[bartol]{N. M. Amin}
\author[marquette]{K. Andeen}
\author[harvard]{C. Arg{\"u}elles}
% ...
\address[brusselslibre]{Universit{\'e} Libre de Bruxelles, Science Faculty CP230, B-1050 Brussels, Belgium}
% ...
\address[harvard]{Department of Physics and Laboratory for Particle Physics and Cosmology, Harvard University, Cambridge, MA 02138, USA}
% ...
\address[loyola]{Department of Physics, Loyola University Chicago, Chicago, IL 60660, USA}
\address[christchurch]{Dept. of Physics and Astronomy, University of Canterbury, Private Bag 4800, Christchurch, New Zealand}
% ...
\address[copenhagen]{Niels Bohr Institute, University of Copenhagen, DK-2100 Copenhagen, Denmark}
\address[dortmund]{Dept. of Physics, TU Dortmund University, D-44221 Dortmund, Germany}
% ...
\address[madisonpac]{Dept. of Physics and Wisconsin IceCube Particle Astrophysics Center, University of Wisconsin{\textemdash}Madison, Madison, WI 53706, USA}
% ...
\address[marquette]{Department of Physics, Marquette University, Milwaukee, WI 53201, USA}
% ...
\address[bartol]{Bartol Research Institute and Dept. of Physics and Astronomy, University of Delaware, Newark, DE 19716, USA}
% ...
\address[zeuthen]{Deutsches Elektronen-Synchrotron DESY, Platanenallee 6, D-15738 Zeuthen, Germany}
\fntext[india]{also at Institute of Physics, Sachivalaya Marg, Sainik School Post, Bhubaneswar 751005, India}
\fntext[chalmers]{also at Department of Space, Earth and Environment, Chalmers University of Technology, 412 96 Gothenburg, Sweden}
\fntext[padovainfn]{also at INFN Padova, I-35131 Padova, Italy}
\fntext[tokyofn]{also at Earthquake Research Institute, University of Tokyo, Bunkyo, Tokyo 113-0032, Japan}
\fntext[padovainfnnow]{now at INFN Padova, I-35131 Padova, Italy}





\title [mode = title]{\texttt{GollumFit}: An IceCube Open-Source Framework for Binned-Likelihood Neutrino Telescope Analyses
{\raisebox{-2.50\depth}{{\href{https://github.com/icecube/GollumFit}{\huge\color{BlueViolet}\faGithub}}}}}

\shorttitle{\texttt{GollumFit}}

\begin{abstract}
We present \texttt{GollumFit}, a framework designed for performing binned-likelihood analyses on neutrino telescope data. \texttt{GollumFit} incorporates model parameters common to any neutrino telescope and also model parameters specific to the IceCube Neutrino Observatory. We provide a high-level overview of its key features and how the code is organized. We then discuss the performance of the fitting in a typical analysis scenario, highlighting the ability to fit over tens of nuisance parameters. We present some examples showing how to use the package for likelihood minimization tasks. This framework uniquely incorporates the particular model parameters necessary for neutrino telescopes, and solves an associated likelihood problem in a time-efficient manner.
\end{abstract}

\begin{keywords}
analysis framework \sep binned likelihood \sep neutrino telescope \sep 
\end{keywords}

\maketitle

\noindent \textbf{PROGRAM SUMMARY}\\
\textit{Program title}: \texttt{GollumFit}\\
\textit{Documentation website}: \href{https://docs.icecube.aq/gollumfit/main/index.html}{https://docs.icecube.aq/gollumfit/main/index.html} \\
\textit{Developer's repository link}: \href{https://github.com/icecube/GollumFit}{https://github.com/icecube/GollumFit} \\
\textit{Licensing provisions}: GNU Lesser General Public License 2.1 (LGPL)\\
\textit{Programming language}: \texttt{C++}, \texttt{Python}\\ 
\textit{Nature of problem}: Statistical analysis of data from neutrino telescope experiments is often complex and computationally demanding, owing to the need to optimize a likelihood function over many parameters that describe sources of systematic uncertainties and quantities of interest.\\
\textit{Solution method}: We introduce a framework that performs binned-likelihood optimization, whose performance can handle the number of parameters typical for a neutrino telescope analysis.
We highlight a method to perform event-by-event reweighting to incorporate the experimental parameters.
In particular, for neutrino telescopes the parameters that incorporate the uncertainties in the atmospheric neutrino flux are common across all experiments and analyses, and are implemented in our framework.
The framework has been designed to be easily extendable in the number of observable dimensions and fit parameters.
Finally, we use an automatic differentiation package to achieve computational speed in the likelihood optimization. \\

\section{Introduction}
\label{sec:intro}

In physics experiments with large quantities of data, analyzing these data to perform physics inferences, whether using a frequentist or Bayesian framework, often uses a likelihood-based approach.
A typical analysis to test hypotheses or extract physical parameters will involve binning the data in some characteristic observable space and comparing it to a corresponding expected distribution computed using Monte Carlo techniques. 
For example, a common situation is where the variation of the Monte Carlo expected distribution is known with respect to a model parameter. 
The best-fit value of this parameter is extracted by maximizing the likelihood over this parameter, which then becomes a commonplace function maximization problem.
For large, complex experiments, as is common in particle physics, this problem quickly faces the curse of dimensionality since there are many model parameters; moreover, the likelihood gradient, which is often required for efficient minimization of the likelihood, may be nontrivial to compute analytically with respect to these many model parameters.
Thus, in the regime of many parameters, efficient and accurate calculation of the likelihood and its gradient is crucial to reduce computation time.  
Neutrino telescopes are one such class of physics experiments that relies on likelihood-based data analyses to extract useful physical results and inferences on parameters.
Neutrino telescopes are typically large, gigaton-scale, volumes of transparent media (such as water or ice) that have been instrumented with photodetectors.
When a neutrino interacts within, or close to, this volume, the resulting secondary particles create Cherenkov light, which can be detected by the photodetectors.
Examples of neutrino telescope experiments include the IceCube Neutrino Observatory~\cite{IceCube:2002eys,IceCube:2006tjp} at the South Pole or the KM3NeT/ARCA detector in the Mediterranean Sea~\cite{KM3Net:2016zxf,KM3NeT:2018wnd}.

Neutrino telescopes often study \textit{diffuse} neutrinos, meaning neutrinos that are roughly isotropic in arrival direction versus originating from a single point source.
Diffuse neutrinos may be atmospheric, meaning they originate from showers caused by cosmic rays impacting the Earth's atmosphere, or they may be astrophysical, meaning they originate from a population of cosmic sources like the Milky Way galaxy, supernovae, or active galactic nuclei. 
A typical diffuse analysis will compare a binned distribution of data with a corresponding distribution computed using Monte Carlo techniques, the latter described by model parameters.
Model parameters unique to a neutrino telescope include the transparency of the medium to light, the efficiency of photodetectors, and the uncertainties on the atmospheric neutrino flux; these all must be incorporated into the likelihood.
A binned likelihood is then maximized over model parameters to quantify the goodness-of-fit of the Monte Carlo distribution to the data distribution; the likelihood fit extracts estimates on model parameters and/or analyzes the preference for an alternate hypothesis. 
Ref.~\cite{Abbasi:2021qfz} is an example of a diffuse IceCube analysis that uses the method outlined above to perform a measurement of the energy spectrum of astrophysical neutrinos. 
Ref.~\cite{IceCube:2024kel} is an IceCube example using the same method but different model parameters to search for sterile neutrinos. 

We present an open-source analysis framework, \texttt{GollumFit}, to perform binned-likelihood analyses of neutrino telescope data.
\texttt{GollumFit} addresses our need to quickly optimize a high-dimensional likelihood over a large and particular set of model parameters.
\texttt{GollumFit} also contains auxiliary helper features that aid in computation speed.
We do not attempt to make \texttt{GollumFit} a general framework for all scientific analyses. 
Instead, we focus upon the shared commonalities of high-energy ($>100\:\text{GeV}$) neutrino telescope analyses: 
\begin{itemize}
    \item the binning of data events in an observable space that typically consists of, but is not limited to, the reconstructed neutrino energy and cosine of the zenith angle (the angle between a neutrino's direction of travel and the local vertical (the ``zenith'') at the detector);
    \item a corresponding set of Monte Carlo simulation events that is parameterized by model parameters, which are experiment- or analysis-specific;
    \item the presence of common classes of model parameters such as detector efficiency and uncertainties in neutrino flux that are universal to all neutrino telescopes.
\end{itemize}
\texttt{GollumFit} constructs the likelihood problem incorporating the data, Monte Carlo, and specific neutrino telescope model parameters.
Moreover, \texttt{GollumFit} includes a set of nominal IceCube Monte Carlo events, parametrized by a collection of model parameters. 
These have been generated from a model of the IceCube detector, which may be used to perform simple fitting to data. 
This set of Monte Carlo events is generated with the LeptonInjector event generator~\cite{IceCube:2020tcq}, and is a subset of the Monte Carlo events used in the analysis of Ref.~\cite{IceCube:2024uzv}.
This software is, to our knowledge, a novel attempt at creating a fast likelihood fitting framework specialized for neutrino telescopes.

\section{Fitting Problem}
\label{sec:fittingproblem}

In this section we will give an overview of the general fitting problem that is encountered by many neutrino telescope analyses, and which \texttt{GollumFit} is designed to solve.
The fitting problem can be summarized as follows: given a binned distribution of data events, and a corresponding binned distribution of Monte Carlo events which depends on model parameters, what are the values of the model parameters such that the Monte Carlo distribution best resembles the data distribution, as quantified by the maximum of a binned likelihood?
In this section, we describe key components of this problem in detail, as applicable to neutrino telescopes: the binning of events, the model parameters that are inferred, and the likelihood to be maximized.

\subsection{Binning}
\label{sec:fittingproblem:binning}

In general, a Monte Carlo event has \textit{true} quantities, $\vec q$, and \textit{reconstructed} quantities, $\vec Q$, where the reconstructed quantities have been subject to the same processing as the data, so that it is directly comparable to what is observed.
In IceCube and other neutrino telescope analyses \cite{IceCube:2024kel,IceCube:2024uzv,IceCube:2016rnb, IceCube:2020phf}, data and Monte Carlo events are often binned in a two-dimensional parameter space. 
The axes of this parameter space are the reconstructed energy, $E\in\vec Q$, of the neutrino and the reconstructed cosine of the zenith, $\cos\theta \in \vec Q$. 
The zenith angle, $\theta$, is defined (for IceCube) as the angle between the neutrino arrival direction and the normal to the Earth's surface at the South Pole; therefore, a value of $\cos\theta=1$ corresponds to a neutrino traveling vertically downwards from overhead at the South Pole; $\cos\theta=0$ corresponds to a neutrino from the horizon regardless of azimuthal direction; $\cos\theta=-1$ corresponds to a neutrino traveling vertically upwards.
\texttt{GollumFit} assumes this two-dimensional binned space by default.
It will create histograms and calculate the likelihood according to this binning. 
A trivial third axis on which events are often binned is the classification of different event morphologies, meaning events that appear as different shaped energy deposits in the detector, owing to different physical processes that create them. 
For instance, in IceCube, \textit{tracks} left by muons can be classified as starting and throughgoing, referring to detected muons that are created within the detector volume (starting) or outside it (throughgoing).
For details on this classification, see Ref.~\cite{IceCube:2024uzv} as an example. 
Figure \ref{fig:expectation_variation} (left) is an example of a histogram of starting tracks that is binned in $E$ and $\cos\theta$.
For simplicity, in this work, we will only discuss morphological categories associated with muon neutrinos, though our code is readily extendable to these other cases.
Other potential morphological categories that could be added are \textit{cascades} and/or \textit{double cascades} as done in Refs.~\cite{IceCube:2020wum,IceCube:2024nhk}.

However, it is clear that not all neutrino telescope analysis situations will have this exact binning. 
One could easily envision another dimension in binning corresponding to, for instance, the right ascension (azimuth) direction of the incoming neutrino, or another binary classifier dimension. 
In these cases, it will be necessary to add a new binning dimension.

\subsection{Model Parameters} \label{sec:parameters}

The model parameters that parametrize the Monte Carlo events of an analysis can, in general, be split into two categories: physics ($\vec{\theta}$) and nuisance ($\vec{\eta}$) parameters. 
The physics parameters are the model parameters of interest which the analysis is designed to infer.
Examples from IceCube include the spectral index of a power law describing a specific component of the neutrino flux, a mixing with a sterile neutrino state that may alter the muon neutrino flux (such as in Ref. \cite{IceCube:2024kel}), or the density of the medium which neutrinos pass through. 
Nuisance parameters are typically the ones which describe the effect of systematic uncertainties on the Monte Carlo events.
They are not the object of an analysis to measure, but still need to be taken into account. 
IceCube of nuisance parameters are the efficiency of the optical detectors, or a scaling factor to capture the uncertainties associated with the atmospheric neutrino flux. 
The distinction between physics and nuisance parameters is purely a conventional one, and is analysis-specific.

In \texttt{GollumFit} we include a collection of nuisance parameters which are common to neutrino telescopes. 
A list of these nuisance parameters is included in table \ref{tab:syst_uncertainties}. 
Below, we list the categories of nuisance parameters that are included, and explain them briefly: 
\begin{itemize}

    \item \textit{Normalization.} \texttt{convNorm} is an overall normalization for the entire neutrino flux.
    
    \item \textit{Conventional flux.} These are nuisance parameters associated with the conventional atmospheric neutrino flux model and its uncertainties. Atmospheric neutrinos are primarily produced by the decay of kaons ($K$) and pions ($\pi$) to muon neutrinos in the atmosphere. 
    \begin{itemize}
        \item $\rho_{\textrm{atm}}$ and $\sigma_{\textrm{K-Air}}$ describe the effect of atmospheric density and meson energy loss, respectively; see Ref. \cite{IceCube:2020tka}. 
        \item The $\textrm{K}$, $\pi$, $\textrm{p}$ and $\textrm{n}$ (10 in total) parameters describe the hadronic yield, as prescribed by the \texttt{DAEMONFLUX} calculation (Ref. \cite{Yanez:2023lsy}) and also used in Ref. \cite{IceCube:2024uzv}.
        \item The $\textrm{GSF}$ parameters (6 in total) describe the cosmic ray spectrum that produces the hadrons associated with conventional flux production, again as prescribed by \texttt{DAEMONFLUX} (Ref. \cite{Yanez:2023lsy}) and again used in Ref. \cite{IceCube:2024uzv}.
    \end{itemize}
    We encode the correlations among the latter 16 \texttt{DAEMONFLUX}-associated parameters in a correlation matrix used during fitting.

    \item \textit{Non-conventional flux.} These parameters are associated with the prompt atmospheric flux and the astrophysical flux, which are each sub-leading compared to the conventional flux, except at high energies. 
    \begin{itemize}
        \item $\Phi^{\textrm{HE}}$, $\Delta\gamma_{1}^{\textrm{HE}}$, $\Delta\gamma_{2}^{\textrm{HE}}$, and $\log_{10}\left(\textrm{E}_{\textrm{break}}^{\textrm{HE}}/\textrm{GeV}\right)$ are the normalization, two spectral indices, and break location to characterize a broken power law astrophysical flux shape. 
        \item \texttt{promptNorm} is the normalization factor for the prompt atmospheric flux component.
        \item $\nu/ \bar{\nu}$ is the neutrino-antineutrino ratio in the astrophysical flux.
    \end{itemize}

    \item \textit{Cross section.} The parameters $\nu\textrm{ Attenuation}$ and $\bar{\nu}\textrm{ Attenuation}$ are factors on the cross section designed to account for any cross section uncertainties; specifically, they parametrize the effect of increasing the neutrino-nucleon cross section on the Earth transparency. 
    This parametrization is general and suitable for any high-energy neutrino telescope.

\end{itemize}
In the following, we list the nuisance parameters that are specific to IceCube. 
For a different telescope, these will likely be different parameters that depend on the specific medium and hardware of that detector. 
\begin{itemize}

    \item \textit{Local detector response.} The parameters $\textrm{DOM eff}$ and $\textrm{Hole Ice}$ characterize the efficiency of the photon detectors (``DOMs'') and the effect of the ice in the boreholes that are optically distinct from the glacial ice. 
    Refer to Ref.~\cite{IceCube:2020tka} for more details. 

     \item \textit{Bulk ice.} These parameters ($\textrm{Ice A}$ and $\textrm{Ice Phs}$, 9 in total) stem from the SnowStorm method (Ref.~\cite{IceCube:2019lxi}) which describes variations in the ice and models them via amplitudes and phases of Fourier modes. As shown in Ref.~\cite{IceCube:2024uzv}, only the first five amplitude and four phase parameters have a large effect on the distribution of events.
    Finally, these 9 parameters are also correlated with a corresponding correlation matrix which should be used when fitting. 
    
\end{itemize}

\begin{table}[h!]
\centering
\begin{tabular}{|l|l|l|}
\hline
\textbf{Parameter} & \textbf{\texttt{GollumFit} Variable} & \textbf{Weighting Method} \\ \hline
\multicolumn{3}{|c|}{Common to all neutrino telescopes} \\ \hline
$\textrm{convNorm}$         & \texttt{convNorm}                & scale factor \\
$\rho_{\textrm{atm}}$       & \texttt{zenithCorrection}        & gradient \\
$\sigma_{\textrm{K-Air}}$   & \texttt{kaonLosses}              & gradient \\
$\textrm{K}_{158G}^{+}$     & \texttt{hadronicHEkp}            & gradient \\
$\textrm{K}_{158G}^{-}$     & \texttt{hadronicHEkm}            & gradient \\
$\pi_{20T}^{+}$            & \texttt{hadronicVHE1pip}         & gradient \\
$\pi_{20T}^{-}$            & \texttt{hadronicVHE1pim}         & gradient \\
$\textrm{K}_{2P}^{+}$       & \texttt{hadronicVHE3kp}          & gradient \\
$\textrm{K}_{2P}^{-}$       & \texttt{hadronicVHE3km}          & gradient \\
$\pi_{2P}^{+}$             & \texttt{hadronicVHE3pip}         & gradient \\
$\pi_{2P}^{-}$             & \texttt{hadronicVHE3pim}         & gradient \\
$\textrm{p}_{2P}$          & \texttt{hadronicVHE3p}           & gradient \\
$\textrm{n}_{2P}$          & \texttt{hadronicVHE3n}           & gradient \\
$\textrm{GSF}_1$           & \texttt{cosmicRay1}              & gradient \\
$\textrm{GSF}_2$           & \texttt{cosmicRay2}              & gradient \\
$\textrm{GSF}_3$           & \texttt{cosmicRay3}              & gradient \\
$\textrm{GSF}_4$           & \texttt{cosmicRay4}              & gradient \\
$\textrm{GSF}_5$           & \texttt{cosmicRay5}              & gradient \\
$\textrm{GSF}_6$           & \texttt{cosmicRay6}              & gradient \\
$\Phi^{\textrm{HE}}/10^{-18}\textrm{GeV}^{-1}\textrm{sr}^{-1}\textrm{s}^{-1}\textrm{cm}^{-2}$ 
                          & \texttt{astroNorm}               & scale factor \\
$\Delta\gamma_{1}^{\textrm{HE}}$ 
                          & \texttt{astroDeltaGamma}         & power law formula \\
$\Delta\gamma_{2}^{\textrm{HE}}$ 
                          & \texttt{astroDeltaGammaSec}      & power law formula \\
$\log_{10}\left(\textrm{E}_{\textrm{break}}^{\textrm{HE}}/\textrm{GeV}\right)$ 
                          & \texttt{astroPivot}              & power law formula \\
$\textrm{promptNorm}$       & \texttt{promptNorm}              & scale factor \\
$\nu/ \bar{\nu}$           & \texttt{NeutrinoAntineutrinoRatio} 
                          & scale factor \\
$\nu\textrm{ Att}$         & \texttt{nuxs}                    & spline \\
$\bar{\nu}\textrm{ Att}$    & \texttt{nubarxs}                 & spline \\
\hline
\multicolumn{3}{|c|}{IceCube-specific Monte Carlo parameters} \\ \hline
$\textrm{DOM}\:{\textrm{eff}}$ & \texttt{domEfficiency}        & spline \\
$\textrm{Hole Ice}$           & \texttt{holeiceForward}         & spline \\
$\textrm{Ice A}_{0}$          & \texttt{icegrad0}               & gradient \\
$\textrm{Ice A}_{1}$          & \texttt{icegrad1}               & gradient \\
$\textrm{Ice A}_{2}$          & \texttt{icegrad2}               & gradient \\
$\textrm{Ice A}_{3}$          & \texttt{icegrad3}               & gradient \\
$\textrm{Ice A}_{4}$          & \texttt{icegrad4}               & gradient \\
$\textrm{Ice Phs}_{1}$        & \texttt{icegrad5}               & gradient \\
$\textrm{Ice Phs}_{2}$        & \texttt{icegrad6}               & gradient \\
$\textrm{Ice Phs}_{3}$        & \texttt{icegrad7}               & gradient \\
$\textrm{Ice Phs}_{4}$        & \texttt{icegrad8}               & gradient \\
\hline
\end{tabular}
\caption{\textit{Table of included nuisance parameters.} We include their conventional name, the variable name in \texttt{GollumFit}, and the method that is used to account for it in reweighting (see section \ref{sec:reweighting}). We have also made a distinction between parameters describing a general neutrino telescope and parameters that are specific to IceCube. For a general description of the nuisance parameters, see section \ref{sec:parameters}. The reweighting method describes exactly how the event weight depends on the nuisance parameter. This can be a gradient, spline, or analytical formula, describing the variation of the weight as a function of the nuisance parameter}
\label{tab:syst_uncertainties}
\end{table}

\subsection{Likelihood}

With any binning scheme, \texttt{GollumFit} calculates a binned likelihood, which compares the binned data and Monte Carlo distributions:
\begin{equation} \label{eq:totalLH}
    \mathcal{L}(\vec\theta, \vec\eta) = \prod_{i \in \{\textrm{bins}\}} \mathcal{L}_{\textrm{eff}}(\mu_i(\vec\theta, \vec\eta), x_i),
\end{equation}
where $\mathcal{L}_{\textrm{eff}}$ is typically a Poisson likelihood.
In \texttt{GollumFit}, $\mathcal{L}_{\textrm{eff}}$ is the effective likelihood obtained in Ref. \cite{Arguelles:2019izp}, which is a modified Poisson likelihood that takes into account the statistical uncertainty in generating Monte Carlo samples.
The variable $i$ iterates over all the bins and $\mu_i$ is the expected value in each bin. 
That is, it is the sum of the weights of all events in the given bin $i$: $\mu_i = \sum_{e\in i} w_e$.
For inference of parameters, we maximize this likelihood to obtain the best fit $\mathcal{L}_{\textrm{BF}}$: 
\begin{equation}
    \mathcal{L}_{\textrm{BF}} = \max_{\vec\theta, \vec\eta}  \mathcal{L}(\vec\theta, \vec\eta) \Pi(\vec\theta, \vec\eta),
\end{equation}
where $\Pi(\vec\theta, \vec\eta)$ represents the penalty terms associated with the priors on the parameters:
\begin{equation}
    \Pi(\vec\theta, \vec\eta) = \prod_{\theta \in \vec\theta} \pi(\theta) \times  \prod_{\eta \in \vec\eta} \pi(\eta).
\end{equation}
The function $\pi$ could be a Gaussian or a uniform function depending on the nature of the prior. 
After obtaining $\mathcal{L}_{\textrm{BF}}$ we will find the corresponding best fit values for $\vec\theta$ and $\vec\eta$. 

In the default version of \texttt{GollumFit}, only nuisance parameters are included, given that the physics parameters for each analysis will be distinct. 
Operationally, any physics parameters are appended to the list of nuisance parameters and treated identically, followed by a simultaneous fit.
\section{The \texttt{GollumFit} Solution}
\label{sec:overview}

In this section we outline the \texttt{GollumFit} framework in the context of the specific aspects of the fitting problem we have outlined in section \ref{sec:fittingproblem}. In particular, we introduce both high- and low-level aspects of the software. We outline the process of reweighting the likelihood, and the specific optimization method used. In a lower-level discussion we outline general usage and explain the FastMC feature.

In this section we make use of publicly-available Monte Carlo events which represents a typical example of what may be used in an analysis; as described in Section~\ref{sec:intro}, this is a set of IceCube Monte Carlo events generated with the LeptonInjector event generator~\cite{IceCube:2020tcq} and a subset what was used most recently in Ref.~\cite{IceCube:2024uzv}. 
They have the true and reconstructed quantities $\vec q$ and $\vec Q$ as described in Section~\ref{sec:fittingproblem:binning}. 
These events are available via the source code repository of \texttt{GollumFit}
In general, any set of Monte Carlo events that conforms to the same format, with quantities $\vec q$ and $\vec Q$, can be used within GollumFit; the supplied set of events is for reference only.


\subsection{Reweighting for Likelihood Evaluation} \label{sec:reweighting}

Each Monte Carlo event, $e$, is associated with a weight, $w_e$, that can readily be modified, or ``reweighted,'' to reflect different physical scenarios without the need to produce a new set of Monte Carlo events.
Statistically, in a given bin, the sum of all weights of events in that bin is the Poisson expectation for the number of observed events.
The weights of Monte Carlo events are parameterized by model parameters; in \texttt{GollumFit}, there are 38 nuisance parameters already included, as described in section \ref{sec:parameters} and table \ref{tab:syst_uncertainties}. 
While Monte Carlo weights depend on $\vec{\theta}$ and $\vec{\eta}$, they may also depend on the event's true and reconstructed quantities ($\vec q_e$ and $\vec Q_e$ respectively), like its zenith and energy, particle type, flux component, etc. 
In \texttt{GollumFit}, non-trivial dependences of $w_e$ on $\vec{\theta}$, $\vec{\eta}$, $\vec Q_e$, or $\vec q_e$, can be parametrized in three main ways:
\begin{enumerate}
    \item analytically, where there is a known formula which relates the change in weight (for instance, a basic scaling factor, or a scaling according to some given function)
    \item by gradient, where the gradient of the weight with respect to a given parameter is known and used to linearly extrapolate the change in weight
    \item by spline, where the change in weight is encoded as the output of a (possibly many-dimensional) spline
\end{enumerate}
The final column in table \ref{tab:syst_uncertainties} lists the exact parametrization used for encoding the dependence $w_e$ on each model parameter (specifically, on $\vec{\eta}$). 
Necessarily, the user will need to determine how to parametrize this dependence of $w$ on any new model parameter they wish to add.

For every event, \texttt{GollumFit} finds the change in $w_e$ every unique set of model parameters, $(\vec{\theta}, \vec{\eta})$; the same is done for every event in the Monte Carlo set.
\texttt{GollumFit} then iterates over every bin and re-computes the reweighted Monte Carlo event distribution. 
Specifically, the expected value in each bin $i$ is computed: $\mu_i = \sum_{e\in i} w_e$.
This reweighted Monte Carlo event distribution is necessary to evaluate the likelihood for a given $(\vec{\theta}, \vec{\eta})$.
The purpose of reweighting is so that the likelihood for any given combination of model parameters can be quickly calculated, instead of re-generating a new set of Monte Carlo events each time, which would be computationally intractable.

\begin{figure*}[htbp]
  \centering
  \includegraphics[width=1.0\textwidth]{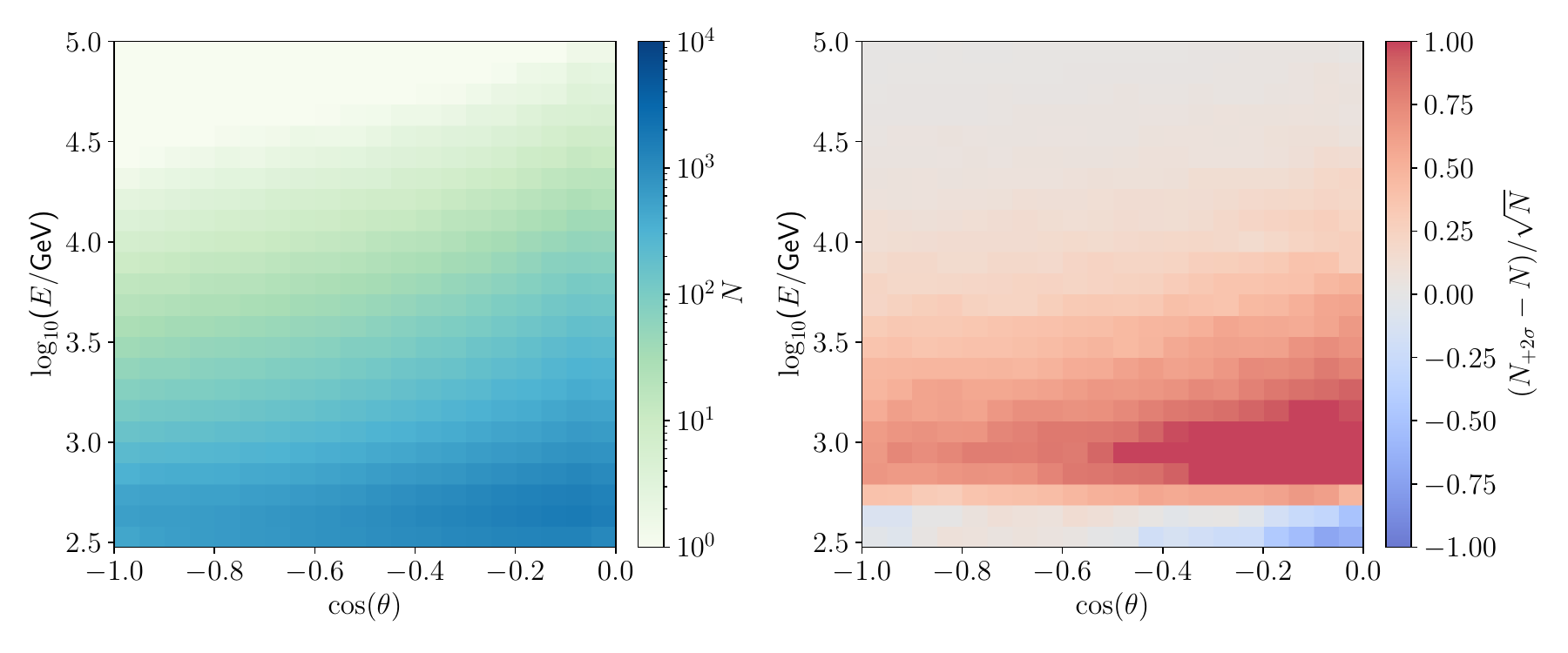}
  \caption{\textit{Left:} a plot of the binned Monte Carlo event distribution, binned in the space of reconstructed log energy ($\log_{10}(E / \text{GeV})$) and the cosine of the reconstructed zenith angle ($\cos\theta$). The color scale indicates the number $N$ of events in each bin. To generate this distribution, all nuisance parameters are held at their prior central value. \textit{Right:} in contrast, a pull plot that compares another Monte Carlo event distribution with the one on the left, assuming the prior central values of all nuisance parameters, with the exception of \texttt{DOMEfficiency}, which has been increased by two times its prior width. The color scale indicates the pull value comparing a new bin count $N_{+2\sigma}$ with $N$. The deficit in the bottom rows of the pull plot is due to the \texttt{DOMEfficiency} shifting lower energy events to higher energies, and these rows contain the lowest energy events in the sample.}
  \label{fig:expectation_variation}
\end{figure*}

As a concrete example of how re-weighting is done, figure \ref{fig:expectation_variation} on the left shows the binned Monte Carlo event distribution, evaluated at the nominal values for all model parameters.
In the same figure on the right, we also show the pull value from comparing another Monte Carlo event distribution which has the parameter \texttt{DOMEfficiency} increased by two times its prior width via reweighting the nominal Monte Carlo distribution.
\texttt{DOMEfficiency} is a nuisance parameter that describes the efficiency of the optical sensors to detect incoming photons.
The horizontal feature, showing a deficit at the bottom of the plot (low energies), occurs due to there being a peak in the spectrum around $\log_{10}(E / \text{GeV})\approx 2.7$. 
As \texttt{DOMEfficiency} increases, events shift from lower to higher energies.
However, the bottom rows do not have even lower energy events that backfill them so they deplete.
In this case, that is the only parameter that has been varied. 
In a realistic fitting scenario, every parameter would be varied simultaneously.

\subsection{Optimization Method}
\texttt{GollumFit}'s reweighting provides a fast and straightforward way to evaluate the likelihood given any location in the model parameter space.
However, the maximization of the posterior $\mathcal{L}\times \Pi$ to find the best fit is a computational challenge, because the dimensionality of $\vec\eta$ is usually nontrivial. 
As mentioned, some IceCube analyses will require about 30 or 40 nuisance parameters. 
A key advantage of \texttt{GollumFit} is the ability to perform this optimization in a short, feasible, timeframe.
\texttt{GollumFit} integrates the public software package, \texttt{PhysTools}, to minimize the likelihood in an efficient way; \texttt{PhysTools} uses the L-BFGS-B algorithm \cite{Zhu:1995lbfgsb1,Zhu:1997lbfgsb2} to perform the likelihood maximization. 

At a low level, this is performed with weighter objects. 
They are \texttt{C++} template classes that are designed to apply a specific correction or transformation to events, based on predefined criteria or models. 
One is defined for every model parameter to be optimized, and this modular construction enables a straightforward way to add additional model parameters to fit over.

\subsection{Usage}

\begin{figure*}[htbp]
  \centering
  \includegraphics[width=0.9\textwidth]{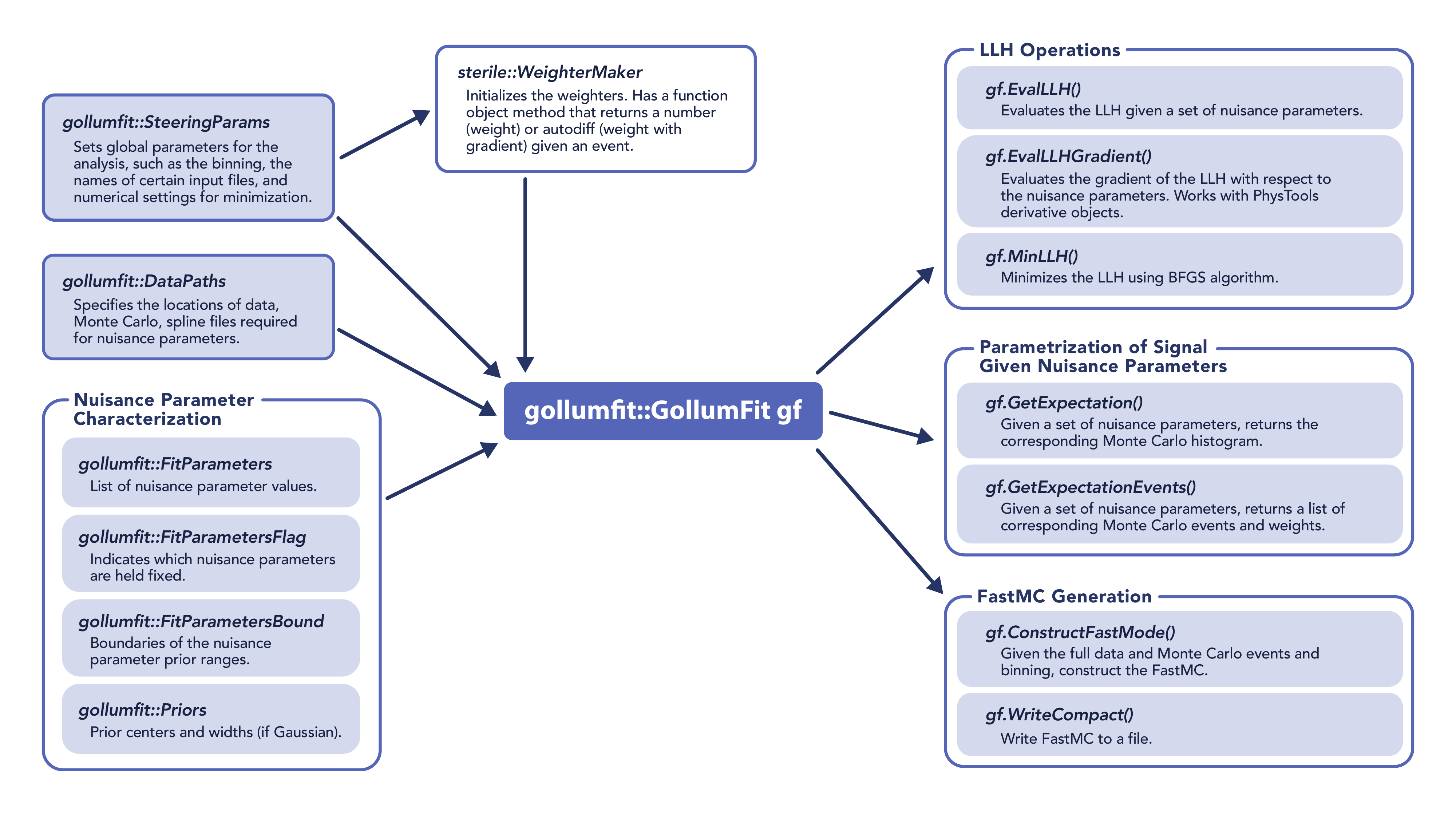}
  \caption{\textit{Overview of GollumFit.} This illustration shows the main components of the main \texttt{GollumFit} object, the necessary inputs, and the most useful functions for analysis purposes.}
  \label{fig:gollumfit_structure}
\end{figure*}

Figure \ref{fig:gollumfit_structure} shows the dependencies of the \texttt{GollumFit} object and the key information that is required to initialize it, and it also shows the key methods that an analysis normally requires.
\texttt{GollumFit} is directly initialized with a \texttt{SteeringParams} and a \texttt{DataPaths} object. 
\texttt{SteeringParams} contains hyperparameters such as the binning scheme in energy and $\cos\theta$, the locations of certain systematics files, and parameters controlling parallel multi-core evaluation of the likelihood.
\texttt{DataPaths} contains the paths to the Monte Carlo, the data, and files for nuisance parameters which rely on gradients or splines for evaluation. 
Nuisance parameters are input by assigning various objects to the value that is desired, for instance, by declaring a list of nuisance parameter priors and assigning them to a \texttt{gollumfit::Priors} object, which is then passed into the main \texttt{GollumFit} object.

The output of figure \ref{fig:gollumfit_structure} can be grouped into three key areas of functionality: likelihood operations, parameterization of the expectation, and FastMC. 

The likelihood methods are the core of the minimization functionality that is central to \texttt{GollumFit}. 
The function \texttt{EvalLLH} returns the negative log-likelihood (LLH) given an input of nuisance parameters, data events, and Monte Carlo events. 
Similarly, \texttt{EvalLLHGradient} returns the gradient of the likelihood with respect to the nuisance parameters. 
Finally, the \texttt{MinLLH} function performs the minimization of the likelihood and returns a \texttt{FitResult} object. 
\texttt{FitResult} includes the best fit values of the nuisance parameters and data like the number of likelihood calculations, the minimum value, and a boolean flag indicating fit failures. 

There are some functions that directly interface with the Monte Carlo event distributions.
\texttt{GetExpectation} and \texttt{GetExpectationEvents} take inputs of nuisance parameter values and respetively return binned histograms and an unbinned weighted vector of events.
These functions are useful when producing plots that aim to show the effect of any changes in parameters on the expected event distributions. 
It can also be helpful in producing the specialized sets of Monte Carlo, generating pseudo-data, performing mismodelling tests, or studying the effect of specific nuisance parameter variations.

Finally, the FastMC functions construct and save the FastMC (see below). 

\subsection{FastMC} \label{sec:fastmc}

\texttt{GollumFit}'s FastMC is a procedure where a substantial Monte Carlo set is consolidated without significantly sacrificing accuracy. 
In this procedure, the expected number of physical events, $N$, is preserved yet the number of consolidated events is guaranteed to be less than or equal to the number of initial Monte Carlo events.
Therefore a procedure to reweight the events, like for \texttt{GollumFit}'s likelihood evaluation, will be significantly sped up (see Fig.~\ref{fig:time_per_LH}) if the consolidated events are used. 
FastMC parses all Monte Carlo events and merges events that are sufficiently close together in the parameter space of true quantities, to produce a new, compact, set of Monte Carlo events.
FastMC's acceleration of the likelihood evaluation depends on the compactness of the resulting MC set.

The FastMC procedure is outlined as follows. 
\begin{enumerate}
    \item A Monte Carlo event, $e$, is described by true quantities, $\vec q_e$, and reconstructed quantities, $\vec Q_e$, and is assigned a weight, $w_e$, which is a function of $\vec q_e$, $\vec Q_e$, and the model (physics $\vec\theta$ \& nuisance $\vec\eta$) parameters: $w_e = w_e(\vec q_e, \vec Q_e, \vec\theta, \vec\eta)$. 
    \item The binning, $B_r$, in the space of reconstructed quantities, is assigned some fixed bin width, $\Delta Q$, for each $Q \in \vec Q$.
    We then define a binning, $B_t$, in the space of true quantities, such that the bin width of the corresponding true value, $q$, is $\Delta q = k\Delta Q$.
    $k$ is a scale parameter assigned a value between 0 and 1, and naturally sets the degree of compression.
    \item We define a \textit{meta bin} as the set of all events which are in the same bin under the binning $B_r$ \textit{and} in the same bin under the binning $B_t$. 
    A \textit{meta event} $\bar{e}$ is an event derived from a given meta bin with properties that are the weighted sum of the constituent events in that meta bin:
    \begin{align}
        \vec q_{\bar{e}} &= \frac{\sum_{e \in \textrm{meta bin}}\vec{q}_e w_e}{\sum_{e \in \textrm{meta bin}}w_e} = \frac{\sum_{e \in \textrm{meta bin}}\vec{q}_e w_e}{w_{\bar{e}}}, \\
        \vec Q_{\bar{e}} &= \frac{\sum_{e \in \textrm{meta bin}}\vec{Q}_e w_e}{\sum_{e \in \textrm{meta bin}}w_e} = \frac{\sum_{e \in \textrm{meta bin}}\vec{Q}_e w_e}{w_{\bar{e}}}.
    \end{align}   
    \item Every event $e$ is then grouped into a meta event $\bar{e}$. 
    These meta events are used as the Monte Carlo events for the analysis. 
\end{enumerate}
Thus in this procedure, we preserve the expected number of physical events ($N = \sum_e w_e = \sum_{\bar{e}}w_{\bar{e}}$) and the number of meta events is less than or equal to the number of initial Monte Carlo events ($\sum_e 1 \geq \sum_{\bar{e}}1$).

There will be a loss of accuracy if there are too few meta events. 
Importantly, $k$ must be selected such that for each bin $i$ under the binning $B_r$, the expected value $\mu_i$ is preserved for all reasonable values of the model parameters:
\begin{equation} \label{eq:optimal_k}
\mu_i = \sum_{e\in i} w_e(\vec q_e, \vec Q_e, \vec\theta, \vec\eta) \approx \sum_{\bar{e} \in i} w_{\bar{e}}(\vec q_{\bar{e}}, \vec Q_{\bar{e}}, \vec\theta, \vec\eta).
\end{equation}
The optimal value of the compression parameter, $k$, should be determined on a case-by-case basis. 
Each analysis will have a different need for accuracy and compression trade-offs, and studies should be done to ensure that the compression is not causing a loss of accuracy in the fit. 
For instance, one way to optimize the value of $k$ is to perform a fit with minimal compression (low $k$) and compare the likelihood and best fit model parameters to those of a fit done with higher compression (higher $k$). 
Nonetheless, the need to find an optimal value of $k$ represents a limitation of the FastMC procedure. 
As of yet, it must be found by hand; a more robust way to determine its value, perhaps by evaluating the validity of Eq.~\ref{eq:optimal_k}, represents a possible avenue for future work.

Within the \texttt{GollumFit} object (see figure \ref{fig:gollumfit_structure}), we include functions that handle FastMC generation and saving.
The function \texttt{ConstructFastMode} performs this FastMC compression procedure internally (i.e. generates meta events) and the function \texttt{WriteCompact} saves the collection of meta events to a file.
When fitting, a path to a FastMC file can be supplied in lieu of normal Monte Carlo files.

\section{Performance}
\label{sec:performance}

We characterize the performance of GollumFit with a series of benchmark tests. 
We use the example Monte Carlo set described in section~\ref{sec:overview} as a reference.

\begin{figure*}[htbp]
  \centering
  \includegraphics[width=1.0\textwidth]{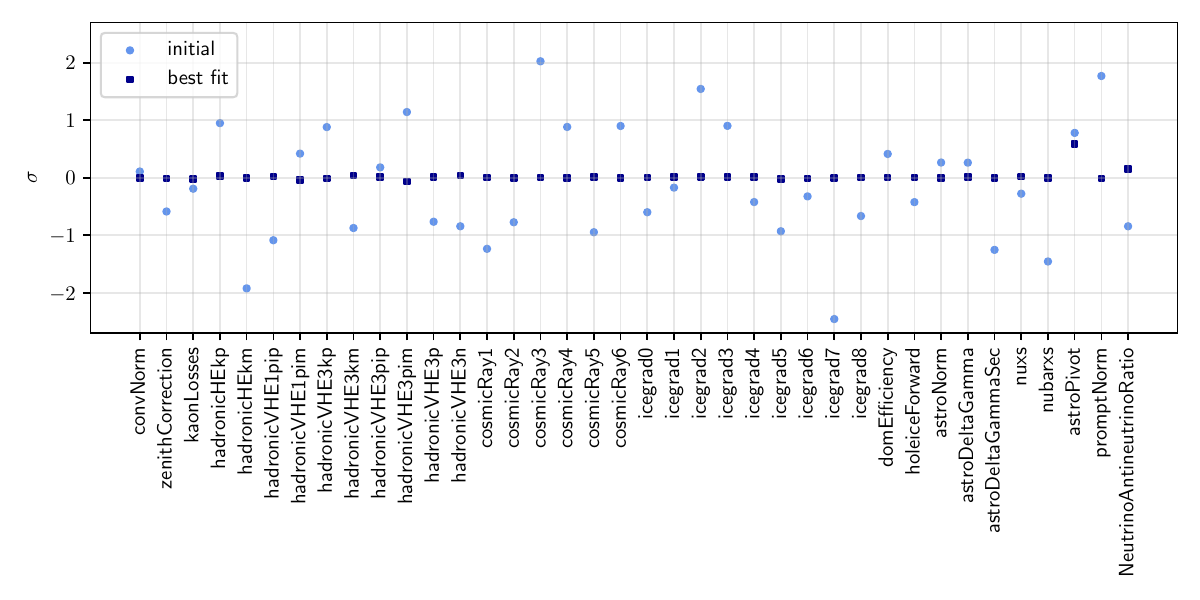}
  \caption{The result of minimization for a single fit, comparing the best-fit and initial value of the nuisance parameters, using $\sigma$, the number of prior standard deviations away from the prior value. The parameter \texttt{astroPivot} has a uniform prior, and no influence on the shape of the flux, and therefore little effect on the likelihood. Hence there is no preferred value to be recovered. A value of $k=0.25$ was used for the FastMC in this fit.}
  \label{fig:fit_to_null}
\end{figure*}

First, to ensure that \texttt{GollumFit} can successfully recover values, we performed a single fit on an output from FastMC assuming nominal values for each nuisance parameter. 
The resulting fit is shown in figure \ref{fig:fit_to_null}, where the vertical axis is showing the location of the fit relative to the prior value in units of prior standard deviations, $\sigma$.
The fit has converged to the nominal (true) value for each nuisance parameter, despite the parameters being initialized randomly away from the true value.
By default, we assume a broken power law in energy to model the astrophysical flux component, and the parameter \texttt{astroPivot} indicates where the break is. 
Given that we assume the same spectral index before and after the break, the value of \texttt{astroPivot} in this example has a uniform prior and no effect on the likelihood, and there is no preferred value to be recovered.

\begin{figure*}[htbp]
  \centering
  \includegraphics[width=0.5\textwidth]{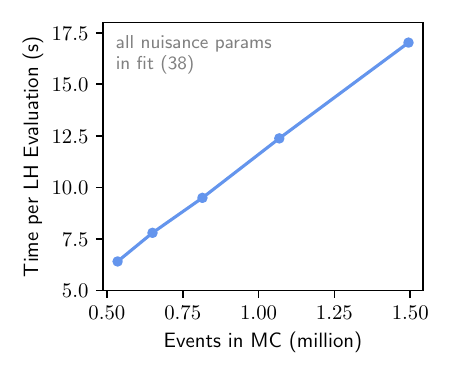}
  \caption{Plot of time per likelihood evaluation as function of the number MC events looped over (corresponding to the size of the FastMC file). For reference, the total MC dataset, before the FastMC compression, consists of 13,061,947 events.}
  \label{fig:time_per_LH}
\end{figure*}

We also compare the time taken to perform one likelihood evaluation. 
We perform the same fit to the nominal set of nuisance parameters given the same random initialization.
The key difference is the size of the Monte Carlo dataset.
We adjust the FastMC scale factor to control the amount of compression and therefore the number of events that are within the Monte Carlo set. 
We time the fit, i.e. the \texttt{MinLLH} function, and divide by the number of likelihood evaluations to get an averaged value for the time taken to perform a single LH evaluation. 
Figure \ref{fig:time_per_LH} shows a plot of the time taken per LH evaluation as a function of the size of the Monte Carlo dataset. 
\texttt{GollumFit} works by looping over each MC event and re-weighting it, so we expect very linear dependence on the size of the Monte Carlo event set.
This is evidenced in figure \ref{fig:time_per_LH}.

\begin{figure*}[htbp]
  \centering
  \includegraphics[width=0.8\textwidth]{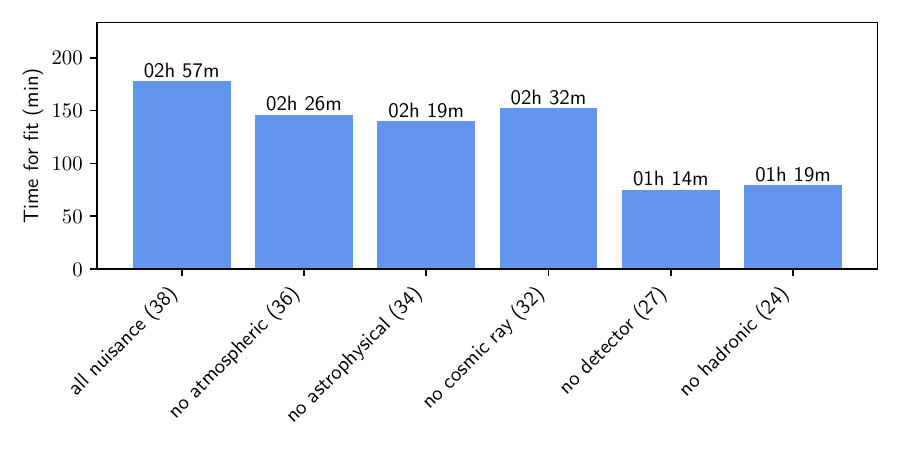}
  \caption{Plot of the time taken for a fit (with identical seed nuisance parameters) to converge. Different bars are for different sets of nuisance parameters turned off. The number of remaining nuisance parameters fitted over is given in the parenthesis. This was performed using 814847 Monte Carlo events, after the FastMC procedure, using a value of $k=0.25$.}
  \label{fig:time_per_param}
\end{figure*}

The time for the fit also depends on the type, and quantity, of nuisance parameters that are being minimized over. 
In figure \ref{fig:time_per_param} we show the total time for performing the minimization, with the different bars representing identically-initialized fits but with different sets of nuisance parameters excluded.
The nominal case, with all nuisance parameters included, takes the longest, but omitting various sets of nuisance parameters offers varying speedups.
Note the number of nuisance parameters is not necessarily the determining factor for the speedup, as the weighting method (see table~\ref{tab:syst_uncertainties}) of the nuisance parameter is also relevant. 
For instance, disabling the cosmic ray parameters removes 6 nuisance parameters, and disabling the astrophysical parameters removes 4 parameters, yet the former case still takes more time. 
Table~\ref{tab:syst_uncertainties} contains the full list of nuisance parameters. 

For all performance tests, \texttt{GollumFit} fitting was run on a system equipped with an Intel® Xeon® Platinum 8480CL processor, featuring 112 cores distributed across 2 sockets with a base frequency of 2.90 GHz and a maximum turbo frequency of 3.80 GHz. 
In our case, only a single core was used.
\section{Conclusions}
\label{sec:conclusions}

We have presented \texttt{GollumFit} as a specialized tool to perform likelihood-based analyses on neutrino telescope data. 
It excels in situations where a multi-dimensional fit is required. 
It can accurately and quickly recover best fit values thanks to its usage of automatic differentiation tools and features such as FastMC. 
Earlier versions of \texttt{GollumFit} have frequently been used in, and have enabled the completion of, various IceCube analyses such as in Refs.~\cite{IceCube:2020phf,IceCube:2024uzv}.
Ultimately, any analysis on a neutrino telescope that relies on a binned-likelihood approach to analyze diffuse neutrino data would likely find \texttt{GollumFit} a useful tool.

We emphasize that \texttt{GollumFit} can be applied to a handful but rapidly growing collection of neutrino telescopes outside of IceCube. 
For instance, only a straightforward extension of the framework is needed for application to KM3NeT~\cite{KM3Net:2016zxf} analyses. 
One would need to use different Monte Carlo, and generate different splines to model the nuisance parameter dependencies, but otherwise the core of the functionality is identical.
One further, exciting, possibility as more neutrino telescopes come online, is the possibility of joint analyses that are performed with more than one neutrino telescope. 
As long as a joint likelihood is defined and different systematic uncertainties are handled appropriately, the core functionality of \texttt{GollumFit} is also suitable for larger scale joint analyses. 
There is no limit to neutrino data only, either. 
Cosmic ray data experiments and analyses, such as the one outlined in Ref.~\cite{CREDO:2020pzy}, or any other measurement requiring a binned likelihood with a similar approach to systematic uncertainties, may benefit from the functionality of \texttt{GollumFit}. 

At the moment, several \texttt{GollumFit} limitations merit consideration. 
As mentioned before, the FastMC compression requires manual tuning of the scale parameter $k$ on a case-by-case basis, and while this provides flexibility, determining the optimal value remains somewhat heuristic. 
For fitting procedures, convergence can be sensitive to the initial parameter values, particularly for analyses with weak constraints on model parameters or when multiple local minima exist in the likelihood surface. 
There exist basic safeguards through parameter bounds and penalty terms, but more sophisticated techniques may improve robustness. 
Fitting stability can be improved, for example, using a judicious choice of Markov Chain Monte Carlo steps, using global minimizers like that of Ref.~\cite{Martinez:2017lzg} in concert with the \texttt{GollumFit} likelihood, or initializing the fit at different points in the model parameter space.
Nontheless, these situational limitations do not invalidate \texttt{GollumFit}'s utility as a robust and efficient tool for the growing neutrino telescope community.

\section*{Acknowledgements}

The IceCube collaboration acknowledges the significant contributions to this manuscript from the University of Texas at Arlington, the Massachusetts Institute of Technology, and Harvard University. 
The authors gratefully acknowledge the support from the following agencies and institutions:
USA {\textendash} U.S. National Science Foundation-Office of Polar Programs,
U.S. National Science Foundation-Physics Division,
U.S. National Science Foundation-EPSCoR,
U.S. National Science Foundation-Office of Advanced Cyberinfrastructure,
Wisconsin Alumni Research Foundation,
Center for High Throughput Computing (CHTC) at the University of Wisconsin{\textendash}Madison,
Open Science Grid (OSG),
Partnership to Advance Throughput Computing (PATh),
Advanced Cyberinfrastructure Coordination Ecosystem: Services {\&} Support (ACCESS),
Frontera and Ranch computing project at the Texas Advanced Computing Center,
U.S. Department of Energy-National Energy Research Scientific Computing Center,
Particle astrophysics research computing center at the University of Maryland,
Institute for Cyber-Enabled Research at Michigan State University,
Astroparticle physics computational facility at Marquette University,
NVIDIA Corporation,
and Google Cloud Platform;
Belgium {\textendash} Funds for Scientific Research (FRS-FNRS and FWO),
FWO Odysseus and Big Science programmes,
and Belgian Federal Science Policy Office (Belspo);
Germany {\textendash} Bundesministerium f{\"u}r Bildung und Forschung (BMBF),
Deutsche Forschungsgemeinschaft (DFG),
Helmholtz Alliance for Astroparticle Physics (HAP),
Initiative and Networking Fund of the Helmholtz Association,
Deutsches Elektronen Synchrotron (DESY),
and High Performance Computing cluster of the RWTH Aachen;
Sweden {\textendash} Swedish Research Council,
Swedish Polar Research Secretariat,
Swedish National Infrastructure for Computing (SNIC),
and Knut and Alice Wallenberg Foundation;
European Union {\textendash} EGI Advanced Computing for research;
Australia {\textendash} Australian Research Council;
Canada {\textendash} Natural Sciences and Engineering Research Council of Canada,
Calcul Qu{\'e}bec, Compute Ontario, Canada Foundation for Innovation, WestGrid, and Digital Research Alliance of Canada;
Denmark {\textendash} Villum Fonden, Carlsberg Foundation, and European Commission;
New Zealand {\textendash} Marsden Fund;
Japan {\textendash} Japan Society for Promotion of Science (JSPS)
and Institute for Global Prominent Research (IGPR) of Chiba University;
Korea {\textendash} National Research Foundation of Korea (NRF);
Switzerland {\textendash} Swiss National Science Foundation (SNSF).






\bibliographystyle{elsarticle-num}
\bibliography{gollumfit}

\begin{thebibliography}{10}
\expandafter\ifx\csname url\endcsname\relax
  \def\url#1{\texttt{#1}}\fi
\expandafter\ifx\csname urlprefix\endcsname\relax\def\urlprefix{URL }\fi
\expandafter\ifx\csname href\endcsname\relax
  \def\href#1#2{#2} \def\path#1{#1}\fi

\bibitem{IceCube:2002eys}
J.~Ahrens, et~al., {Icecube - the Next Generation Neutrino Telescope at the South Pole}, Nucl. Phys. B Proc. Suppl. 118 (2003) 388--395.
\newblock \href {http://arxiv.org/abs/astro-ph/0209556} {\path{arXiv:astro-ph/0209556}}, \href {https://doi.org/10.1016/S0920-5632(03)01337-9} {\path{doi:10.1016/S0920-5632(03)01337-9}}.

\bibitem{IceCube:2006tjp}
A.~Achterberg, et~al., {First Year Performance of The IceCube Neutrino Telescope}, Astropart. Phys. 26 (2006) 155--173.
\newblock \href {http://arxiv.org/abs/astro-ph/0604450} {\path{arXiv:astro-ph/0604450}}, \href {https://doi.org/10.1016/j.astropartphys.2006.06.007} {\path{doi:10.1016/j.astropartphys.2006.06.007}}.

\bibitem{KM3Net:2016zxf}
S.~Adrian-Martinez, et~al., {Letter of intent for KM3NeT 2.0}, J. Phys. G 43~(8) (2016) 084001.
\newblock \href {http://arxiv.org/abs/1601.07459} {\path{arXiv:1601.07459}}, \href {https://doi.org/10.1088/0954-3899/43/8/084001} {\path{doi:10.1088/0954-3899/43/8/084001}}.

\bibitem{KM3NeT:2018wnd}
S.~Aiello, et~al., {Sensitivity of the KM3NeT/ARCA neutrino telescope to point-like neutrino sources}, Astropart. Phys. 111 (2019) 100--110.
\newblock \href {http://arxiv.org/abs/1810.08499} {\path{arXiv:1810.08499}}, \href {https://doi.org/10.1016/j.astropartphys.2019.04.002} {\path{doi:10.1016/j.astropartphys.2019.04.002}}.

\bibitem{Abbasi:2021qfz}
R.~Abbasi, et~al., {Improved Characterization of the Astrophysical Muon\textendash{}neutrino Flux with 9.5 Years of IceCube Data}, Astrophys. J. 928~(1) (2022) 50.
\newblock \href {http://arxiv.org/abs/2111.10299} {\path{arXiv:2111.10299}}, \href {https://doi.org/10.3847/1538-4357/ac4d29} {\path{doi:10.3847/1538-4357/ac4d29}}.

\bibitem{IceCube:2024kel}
R.~Abbasi, et~al., {A search for an eV-scale sterile neutrino using improved high-energy $\nu_\mu$ event reconstruction in IceCube} (5 2024).
\newblock \href {http://arxiv.org/abs/2405.08070} {\path{arXiv:2405.08070}}.

\bibitem{IceCube:2020tcq}
R.~Abbasi, et~al., {LeptonInjector and LeptonWeighter: A neutrino event generator and weighter for neutrino observatories}, Comput. Phys. Commun. 266 (2021) 108018.
\newblock \href {http://arxiv.org/abs/2012.10449} {\path{arXiv:2012.10449}}, \href {https://doi.org/10.1016/j.cpc.2021.108018} {\path{doi:10.1016/j.cpc.2021.108018}}.

\bibitem{IceCube:2024uzv}
R.~Abbasi, et~al., {Methods and stability tests associated with the sterile neutrino search using improved high-energy $\nu_\mu$ event reconstruction in IceCube} (5 2024).
\newblock \href {http://arxiv.org/abs/2405.08077} {\path{arXiv:2405.08077}}.

\bibitem{IceCube:2016rnb}
M.~G. Aartsen, et~al., {Searches for Sterile Neutrinos with the IceCube Detector}, Phys. Rev. Lett. 117~(7) (2016) 071801.
\newblock \href {http://arxiv.org/abs/1605.01990} {\path{arXiv:1605.01990}}, \href {https://doi.org/10.1103/PhysRevLett.117.071801} {\path{doi:10.1103/PhysRevLett.117.071801}}.

\bibitem{IceCube:2020phf}
M.~G. Aartsen, et~al., {eV-Scale Sterile Neutrino Search Using Eight Years of Atmospheric Muon Neutrino Data from the IceCube Neutrino Observatory}, Phys. Rev. Lett. 125~(14) (2020) 141801.
\newblock \href {http://arxiv.org/abs/2005.12942} {\path{arXiv:2005.12942}}, \href {https://doi.org/10.1103/PhysRevLett.125.141801} {\path{doi:10.1103/PhysRevLett.125.141801}}.

\bibitem{IceCube:2020wum}
R.~Abbasi, et~al., {The IceCube high-energy starting event sample: Description and flux characterization with 7.5 years of data}, Phys. Rev. D 104 (2021) 022002.
\newblock \href {http://arxiv.org/abs/2011.03545} {\path{arXiv:2011.03545}}, \href {https://doi.org/10.1103/PhysRevD.104.022002} {\path{doi:10.1103/PhysRevD.104.022002}}.

\bibitem{IceCube:2024nhk}
R.~Abbasi, et~al., {Observation of Seven Astrophysical Tau Neutrino Candidates with IceCube}, Phys. Rev. Lett. 132~(15) (2024) 151001.
\newblock \href {http://arxiv.org/abs/2403.02516} {\path{arXiv:2403.02516}}, \href {https://doi.org/10.1103/PhysRevLett.132.151001} {\path{doi:10.1103/PhysRevLett.132.151001}}.

\bibitem{IceCube:2020tka}
M.~G. Aartsen, et~al., {Searching for eV-scale sterile neutrinos with eight years of atmospheric neutrinos at the IceCube Neutrino Telescope}, Phys. Rev. D 102~(5) (2020) 052009.
\newblock \href {http://arxiv.org/abs/2005.12943} {\path{arXiv:2005.12943}}, \href {https://doi.org/10.1103/PhysRevD.102.052009} {\path{doi:10.1103/PhysRevD.102.052009}}.

\bibitem{Yanez:2023lsy}
J.~P. Ya\~nez, A.~Fedynitch, {Data-driven muon-calibrated neutrino flux}, Phys. Rev. D 107~(12) (2023) 123037.
\newblock \href {http://arxiv.org/abs/2303.00022} {\path{arXiv:2303.00022}}, \href {https://doi.org/10.1103/PhysRevD.107.123037} {\path{doi:10.1103/PhysRevD.107.123037}}.

\bibitem{IceCube:2019lxi}
M.~G. Aartsen, et~al., {Efficient propagation of systematic uncertainties from calibration to analysis with the SnowStorm method in IceCube}, JCAP 10 (2019) 048.
\newblock \href {http://arxiv.org/abs/1909.01530} {\path{arXiv:1909.01530}}, \href {https://doi.org/10.1088/1475-7516/2019/10/048} {\path{doi:10.1088/1475-7516/2019/10/048}}.

\bibitem{Arguelles:2019izp}
C.~A. Arg\"uelles, A.~Schneider, T.~Yuan, {A binned likelihood for stochastic models}, JHEP 06 (2019) 030.
\newblock \href {http://arxiv.org/abs/1901.04645} {\path{arXiv:1901.04645}}, \href {https://doi.org/10.1007/JHEP06(2019)030} {\path{doi:10.1007/JHEP06(2019)030}}.

\bibitem{Zhu:1995lbfgsb1}
R.~H. Byrd, P.~Lu, J.~Nocedal, C.~Zhu, A limited memory algorithm for bound constrained optimization, SIAM Journal on Scientific Computing 16~(5) (1995) 1190--1208.
\newblock \href {https://doi.org/10.1137/0916069} {\path{doi:10.1137/0916069}}.

\bibitem{Zhu:1997lbfgsb2}
C.~Zhu, R.~H. Byrd, P.~Lu, J.~Nocedal, Algorithm 778: L-bfgs-b: Fortran subroutines for large-scale bound-constrained optimization, ACM Trans. Math. Softw. 23~(4) (1997) 550–560.
\newblock \href {https://doi.org/10.1145/279232.279236} {\path{doi:10.1145/279232.279236}}.

\bibitem{CREDO:2020pzy}
P.~Homola, et~al., {Cosmic Ray Extremely Distributed Observatory}, Symmetry 12~(11) (2020) 1835.
\newblock \href {http://arxiv.org/abs/2010.08351} {\path{arXiv:2010.08351}}, \href {https://doi.org/10.3390/sym12111835} {\path{doi:10.3390/sym12111835}}.

\bibitem{Martinez:2017lzg}
G.~D. Martinez, J.~McKay, B.~Farmer, P.~Scott, E.~Roebber, A.~Putze, J.~Conrad, {Comparison of statistical sampling methods with ScannerBit, the GAMBIT scanning module}, Eur. Phys. J. C 77~(11) (2017) 761.
\newblock \href {http://arxiv.org/abs/1705.07959} {\path{arXiv:1705.07959}}, \href {https://doi.org/10.1140/epjc/s10052-017-5274-y} {\path{doi:10.1140/epjc/s10052-017-5274-y}}.

\end{thebibliography}

\pagebreak

\end{document}